\documentclass[a4paper,12pt]{amsart}

\usepackage{graphicx}

\theoremstyle{plain}

\theoremstyle{definition}

\def\hd{HD\,}
\def\l{$\lambda$}
\def\ha{H$\alpha$}
\def\hii{H\,{\sc ii}}
\def\hei{He\,{\sc i}}
\def\heii{He\,{\sc ii}}
\def\nii{N\,{\sc{iii}}}
\def\niii{N\,{\sc{iii}}}

\def\s{S\,{\sc{iv}}}
\def\ciii{C\,{\sc{iii}}}

\def\kms{km\,s$^{-1}$}

\def\rsol{R$_{\odot}$}
\def\msol{M$_{\odot}$}
\def\xmm{{\sc XMM}\emph{-Newton}}

\def\ch{\emph{{\sc Chandra}}}

\title{Spectroscopy of Massive Stars$^a$\footnote{$^a$Based on lectures given at Padova University in December 2005.}}

 \author[Y. Naz\'e]{Ya\"el Naz\'e$^b$}
 \thanks{$^b$Postdoctoral Researcher F.N.R.S. (Belgium)}
 \address[Y. Naz\'e]{\newline
 Ya\"el Naz\'e\newline
 Institut d'Astrophysique et de G\'eophysique, Universit\'e de Li\`ege, Belgium
 }
 \email{naze@astro.ulg.ac.be}

 \date{}

\begin{document}
\maketitle

\begin{abstract}
Although rare, massive stars, being the main sources of ionizing radiation, chemical enrichment and mechanical energy in the Galaxy, are the most important objects of the stellar population. This review presents the many different aspects of the main tool used to study these stars, i.e. spectroscopy. The first part consists in an introduction on these objects and their physical properties (mass, wind, evolution, relation with their environment). Next, the spectral behaviour of single massive stars is investigated, in the visible as well as in the X-ray domain. Finally, the last part of this paper deals with massive binaries, especially those exhibiting a colliding wind phenomenon.
\end{abstract}

% ===== start your paper here ======
\section{Astrophysics of Massive stars: Introduction}

\subsection{Main characteristics}
A star is considered ``massive" if it can ignite carbon burning in its core during the late stages of its evolution. Such stars are the progenitors of black holes and neutron stars and have masses larger than 10\msol. Since they go further into nucleosynthesis stages than stars like the Sun, these stars are the most important sources of chemical enrichment in galaxies. Such massive stars, of spectral type O, are blue and bright objects. Their luminosities amount to $10^5-10^6$L$_{\odot}$: such objects are thus visible from far away in the Universe. In addition, their effective temperatures are larger than 30kK, meaning that the majority of their radiation is emitted in the ultraviolet (UV). Massive stars are therefore the main sources of ionizing radiation in galaxies, and this explains why these stars are surrounded by bright nebulae that are \hii\ regions of ionized gas. \\

The distribution of mass amongst stars follows a law of the form $dN = K M^{-\alpha}dM=\Gamma$ with $N$ the number of stars in the mass interval $[m,m+dm]$, $K$ a constant and $M$ the initial mass of stars. The parameter $\alpha$ is generally considered to be 2.35 (the Salpeter value). Consequently, the number of stars decreases when the mass increases. Massive stars are therefore very rare among the stellar population: for each star with a mass between 60 and 120\,M$_{\odot}$ there form 250 stars with masses between 1 and 2\,M$_{\odot}$. This paucity also means that such stars are also generally distant: the nearest O-type star, $\zeta$ Oph, is situated between 417 and 509 light years; whereas the nearest Wolf-Rayet star (see below) belongs to the binary system $\gamma^2$ Vel and is somewhere between 740 and 975 light years from the Earth.\\

Whilst the lowest possible mass of a massive star is generally well known, the question of the maximum mass of these stars is not settled yet, mainly because of the difficulties in measuring the stellar masses. Estimating the mass of an object can be done by modelling its spectrum, but these models are not always reliable, especially when approaching the limits of the parameter space. The use of a mass-luminosity relation has also been widely popular amongst astronomers but it can result in unrealistic large masses when the object is not spatially resolved. For example, R136a, at the core of the 30 Doradus nebula in the LMC, was once thought to be a single, 1000-2500\msol\ star. It was later found that R136a was actually composed of a dozen components (Weigelt et al. 1991, see also Fig.\ \ref{r136}). In fact, the only reliable method for deriving masses is to observe eclipsing binaries. Using Kepler's laws, one can then constrain all physical parameters of the system by observing the photometric eclipses and the spectroscopic signature of the orbital motion. However, we may note that only young binaries, where no interactions have taken place, can lead to reliable, typical masses. The most massive stars detected so far by this method belong to WR20a and have masses of 82 and 83 \msol\ (Rauw et al. 2005, see Fig.\ \ref{wr20a}). In addition, another method has been used by Figer (2005) to derive an upper limit on the stellar mass: when observing the Arches cluster, he should have detected 20 to 30 stars of mass larger than 150\msol, but this was not the case (see Fig.\ \ref{arches}). He therefore concludes that there exists no star with a mass larger than 150\msol. We emphasize that this upper limit is much larger than the actual largest mass observed in WR20a. \\

  \begin{figure}
  \centering
  \caption{R136a, the cluster at the core of 30 Doradus in the Large Magellanic Cloud (LMC), was once thought to be a single, supermassive star. \copyright\ HST}
  \label{r136}
  \end{figure}

  \begin{figure}
  \centering
  \caption{WR20a, the most massive system ever weighed, is an eclipsing binary ({\it Top}) and a spectroscopic binary ($Bottom$): the masses of the components can thus be evaluated precisely. (G. Rauw, private communication)}
  \label{wr20a}
  \end{figure}

  \begin{figure}
  \centering
  \caption{{\it Top:} The Arches cluster, observed here by HST, contains more than 2000 stars. $Bottom:$ In this cluster, the existence of many stars with more than 150 \msol\ had been predicted theoretically but none was found. \copyright\ HST}
  \label{arches}
  \end{figure}

It can also be noted that the very first generation of massive stars (when there was no metal\footnote{Note that in astrophysics, all elements heavier than helium are called `metals'.} in the Universe, i.e. population III objects), could have reached much larger masses (hundreds to 1000\msol). They could have given birth to intermediate-mass black holes, that might have become seeds for the supermassive black holes at the center of galaxies. These first stars should also have been very luminous and are therefore thought to be responsible for the re-ionisation of the Universe approximately one billion years after the Big Bang. Finally, they would have been the first celestial objects to build metals, sowing the whole Universe. However, all this is still very putative.\\

\subsection{Classification of Massive Stars}
There exist two ways of classifying O-type stars: the spectral morphology or the determination of line ratios. 

The first criterion fits the general philosophy of the MKK classification scheme, where a sample of `typical' standard spectra is constructed and serves as comparison when determining spectral classes. For O-type stars, the main tool of this kind is the atlas made by Walborn \& Fitzpatrick (1990), which provides low-resolution spectra in the 4000-4700 \AA\ range (see Fig.\ \ref{walb}). As O-type stars are very hot, the lines useful for classification are indeed helium lines: the stronger the \heii\ lines compared to the \hei\ lines, the earlier (i.e. the hotter) the star. The transition from O to B-type stars occurs when the \heii\l4542 line is extremely weak, barely detectable. Note that a morphological classification is also used to classify evolved massive stars, i.e. Wolf-Rayet stars.\\

  \begin{figure}
  \centering
  \caption{Excerpt from the digital atlas of Walborn \& Fitzpatrick (1990).}
  \label{walb}
  \end{figure}

\begin{table}
\caption{ Classification criteria for O-type stars using EW measurements (van der Hucht 1996): $\log W'=\log EW_{4471}-\log EW_{4542}$; $\log W''=\log EW_{4088}-\log EW_{4143}$; and $\log W'''=\log EW_{4388}+\log EW_{4686}$. Note that the recently defined spectral types O2 and O3.5 (Walborn et al. 2002) are not presented here.
\label{spectraltype}} 
\small
\begin{center} 
\begin{tabular}{l c | l c || l c}
\hline 
Subtype & $\log W'$ & Subtype & $\log W'$ & Subclass & \\
\hline\hline
O3  & $<-0.9$           & O7.5 & 0 to 0.09   & I  & $\log W''>0.3$\\
O4  & $-0.9$ to $-0.61$ & O8   & 0.1 to 0.19 & III& $0.11<\log W''<0.3$\\
O5  & $-0.6$ to $-0.46$ & O8.5 & 0.2 to 0.29 & V  & $-0.2<\log W''<0.1$\\
O5.5& $-0.45$ to $-0.31$& O9   & 0.3 to 0.44 & I  & $\log W'''<5.15$\\
O6  & $-0.3$ to $-0.21$ & O9.5 & 0.45 to 0.65& III& $5.15<\log W'''<5.4$\\
O6.5& $-0.2$ to $-0.11$ & O9.7 & 0.66 to 1.  & V  & $\log W'''>5.4$\\
O7  & $-0.1$ to $-0.01$ &      &             &    & \\
\hline 
\end{tabular} 
\end{center} 
\end{table} 

The second criterion is more quantitative and uses the determination of equivalent widths (EWs). The EW of a specific line represents the width of a rectangular line of the same area as the actual observed line. Conti \& Frost (1977) and Mathys (1988,1989) have shown that the spectral type of an O star can be determined by comparing the EW of the \hei\l4471 line to that of the \heii\l4542 line (see Table \ref{spectraltype}). Attempts have been made to create a similar quantitative scale in other parts of the electromagnetic spectrum but the Conti-Mathys scale is still the most popular one. The luminosity class can also be deduced from similar EW measurements (see the same Table).\\

\subsection{Stellar Winds}
It is well known that the Sun possesses a solar wind, which is notably responsible for generating polar auroras on Earth. The origin of this wind is linked to the existence of a solar corona. Outside the photosphere ($\sim$6000K), the temperature of the gas increases to reach $10^6$K in the corona. At such temperatures, the gas pressure is very high and it then expands naturally in the lower pressure surrounding regions. By this mechanism, the Sun loses $\sim10^{-14}$M$_{\odot}$yr$^{-1}$.\\

In comparison, massive stars lose $10^{-7}-10^{-4}$M$_{\odot}$yr$^{-1}$ and they do not possess any corona. In fact, the presence of winds in luminous stars is linked to the large luminosity of these objects. In the atmosphere of such stars, there is a transfer of momentum from the photons of the stellar radiation field to the material surrounding the star. In other words, the winds are driven by the absorption of the stellar radiation and its subsequent scattering (see e.g. Lamers \& Cassinelli 1999). If we consider an atom moving in a radial direction that absorbs a photon of frequency $\nu$, its momentum becomes \\
\hspace*{2cm} $mv'_r=mv_r+\frac{h\nu}{c}$. \\
A moment later, the atom re-emits the photon at an angle $\alpha$ (see Fig.\ \ref{windscat}) and the momentum along the radial direction becomes\\ 
\hspace*{2cm} $mv''_r=mv'_r-\frac{h\nu'}{c}\cos(\alpha)$. \\

  \begin{figure}
  \caption{The stellar wind of massive stars is line-driven. Atoms and ions absorb and then re-emit the stellar radiation. }
  \label{windscat}
  \end{figure}

  \begin{figure}
  \centering
  \caption{Formation of a P Cygni profile, resulting from the superposition of an emission profile and an absorption profile. The emission comes from light scattered into the line-of-sight from all regions of the wind (front, blueshifted wind and back, redshifted wind). The absorption comes from light scattered away the line-of-sight by atoms between the star and the observer: in this part, the wind is coming towards us, and the absorption is thus blueshifted. }
  \label{pcyg}
  \end{figure}

The atom will mainly absorb photons at the frequency of specific lines. The frequency, in the rest frame, of such a line will be denoted $\nu_0$. If the atmosphere were static, only radiation near the photosphere would be absorbed. However, since there is a velocity gradient in the wind, each region actually absorbs at a different frequency because of Doppler shift. If the atom is moving with a velocity $v_r$ relative to the star, source of photons, it will see stellar photons with a shifted frequency $\nu(1-v_r/c)$, so the absorbed photon has a frequency $\nu=\nu_0(1+v_r/c)$ in the rest frame of the star. When the photon is subsequently re-emitted, an observer will see that an atom with a velocity of $v'_r$ has emitted a photon of frequency $\nu'=\nu_0(1+v'_r/c)$. Therefore, the final velocity is\\ 
\hspace*{2cm} $v''_r=v_r+\frac{h\nu_0}{mc}(1+v_r/c)-\frac{h\nu_0}{mc}(1+v'_r/c)cos(\alpha)$.\\

Replacing $v'_r$ by the value derived above and assuming that $v<<c$ and $h\nu_0<<mc^2$, we then find: \\
\hspace*{2cm} $v''_r-v_r=\frac{h\nu_0}{mc}(1-cos(\alpha))$. \\
Therefore, we can see that forward scattering ($\alpha$=0) does not increase the momentum of the atom whereas backward scattering ($\alpha$=180$^{\circ}$) increases it by $2\frac{h\nu_0}{c}$. Since the scattering takes place randomly, we can calculate the average momentum gained by averaging over a solid angle of $4\pi$ : \\ 
\hspace*{2cm} $< \Delta mv>=\frac{h\nu_0}{c}\frac{1}{4\pi}\int_0^{\pi}(1-cos(\alpha))2\pi \sin(\alpha) d\alpha=\frac{h\nu_0}{c}$.\\

If the photons were coming from any direction, the net momentum gain would be zero. However, since they come only from the star, a radial acceleration appears, which results in a so-called beta velocity law: \\
\hspace*{2cm} $v(r)= v_{\infty}(1-\frac{R_*}{r})^{\beta}$ \\
where $R_*$ is the stellar radius and $\beta$ amounts to 0.8-1 for O-type stars and 1-2 for WR stars.  The terminal velocities $v_{\infty}$ can reach 1000-3000 \kms\ for massive stars, whereas the velocity of the solar wind is much lower, only 400-700 \kms. \\

The acceleration is mainly provided by the absorption and re-emission of UV photons in the lines of the most abundant metallic ions (CNO+Fe). As a consequence, acceleration is less effective and the mass-loss rate $\dot M$ is much smaller in a low metallicity environment. The energy gained by the metallic ions is subsequently shared with all other particles through interactions with them (Coulomb coupling), so that the whole wind expands. If all photons leaving the stars were absorbed or scattered, the momentum of the wind would equal that of the radiation, i.e. $\dot M_{max}v_{\infty}=L/c$ - this is called the single scattering limit. However, note that a photon can be absorbed and re-emitted several times! \\

The presence of the wind affects the emergent spectrum. The wind acceleration, especially in the UV, generates P Cygni line profiles (see Fig.\ \ref{pcyg}). In addition, some emission lines (e.g. \niii\l\l4634,4641 or \heii\l4686) will be formed in the wind, and not in the photosphere as most of the absorption lines. The spectra of Wolf-Rayet stars give an extreme example of this: they display exclusively broad emission lines formed in their thick stellar wind (see Fig.\ \ref{wr}). \\

  \begin{figure}
  \centering
  \caption{Examples of spectra of Wolf-Rayet stars. (G. Rauw, private communication)}
  \label{wr}
  \end{figure}

The radiation-driven process is an unstable one, so winds of massive stars are intrinsically unstable. Therefore, the wind is more patchy than a smooth ejection of matter. The mass-loss diagnostic lines, like \ha, thus arise in small, dense regions (called clumps): if the lines are analyzed with a spherically symmetric smooth wind model,  the mass-loss rate will be overestimated. Generally, the filling factor of the wind, i.e. the ratio between the volume of the clumps and the total volume, is found to be 0.05--0.25.\\

Since the winds are driven by radiation, the mechanical momentum of the wind should be related to the luminosity of the star. This relation is often expressed as the ``wind-momentum luminosity relation":\\
\hspace*{2cm} $\log \left(\dot M v_{\infty} \left[ \frac{R_*}{R_{\odot}} \right]^{0.5} \right) = \log D_0 + x \log (L_*/L_{\odot})$\\
where $D_0$ is a function of the spectral type and the luminosity class of the star and $x$ depends on the spectral type, luminosity class, and metallicity of the star (see Herrero 2005 and references therein).\\

\subsection{Evolution}
Although massive stars have been observed for centuries, astronomers still debate on how they actually form. In fact, these stars can ignite hydrogen core burning before they have reached their final mass: the radiative pressure of the light then emitted should prevent any further direct accretion. Three main theories have been proposed to overcome this problem. Massive stars could form through the coalescence of lower mass proto-stars, through accretion from a circumstellar disk (just as low-mass stars, but with a higher accretion rate), or through competitive accretion (the most massive clusters producing the most massive stars). Up to now, several evidences seem to favor the accretion model (e.g. Chini et al. 2004, see Fig.\ \ref{accret}). Note that the formation of massive binaries is not completely understood either: tidal capture and fragmentation of the unstable accretion disk have been proposed to explain their existence.\\

  \begin{figure}
  \centering
  \caption{In M17, a baby massive star accretes matter from a circumstellar disk (here seen in silhouette). \copyright\ ESO}
  \label{accret}
  \end{figure}

Once they are formed, massive stars emit large quantities of radiation: they actually burn the hydrogen in their core at a much higher rate of nuclear reactions than lower mass objects. Therefore, although massive stars have more  ``fuel'', they have a much shorter lifetime: a few million years, whereas the Sun will live ten billion years. Because of this short life, these stars do not generally have the time to wander far away from their birth place. Most of the massive stars are thus found in clusters, and the few isolated objects are generally considered as having formed in a cluster and been ejected later on, because of tidal interactions or following a supernova kick (if the star belonged to a binary system).\\

If the formation of massive stars is not completely understood, this is also the case for the details of their subsequent evolution and death. The most popular evolution scenario is the so-called Conti-scenario: massive stars follow the path O$\rightarrow$X$\rightarrow$H-poor WN$\rightarrow$WC. WN and WC are two types of Wolf-Rayet (WR) stars, WN being nitrogen enriched and WC carbon-enriched; X is a phase that depends on the mass of the star, i.e. a red supergiant (RSG) phase if the star has $M<40$\msol\ and a luminous blue variable (LBV) stage otherwise (for more details, see Maeder et al. 2005 and references therein). The wind already plays an important role during the main-sequence lifetime since it is responsible for steadily decreasing the mass of the star at a rate of about $10^{-7}-10^{-6}$M$_{\odot}$yr$^{-1}$. However, the mass-loss further increases during the later stages of the evolution of massive stars: the mass-loss rate of typical Wolf-Rayet stars is $10^{-5}-10^{-4}$M$_{\odot}$yr$^{-1}$ and gigantic mass ejections (whose actual trigger is not known) can even take place during the RSG or LBV phases (see Fig.\ \ref{eta}). The separation between the photosphere and the very thick wind is thus less and less clear as the star evolves and material that was once in the convective core of the star, where the nuclear reactions happen, becomes exposed at the stellar surface. This explains the anomalous abundances observed in the spectrum of these stars: for example, WN stars show at the surface the results of hydrogen burning through the CNO process whereas WC stars expose layers that, since they were originally deeper inside the star, have been rather affected by helium burning. Finally, it is supposed that the massive stars end their life in a supernova explosion. Massive progenitors have even been proposed for the powerful gamma-ray bursts.

  \begin{figure}
  \centering
  \caption{{\it Left:} $\eta$ Carinae is an LBV that underwent two eruptions in the 19th century, losing 1-2\msol\ in the process. $Right:$ WR124, a Wolf-Rayet star, is surrounded by thick ejecta from previous evolutionary stages. \copyright\ HST}
  \label{eta}
  \end{figure}

\subsection{Interactions with the surroundings}

Stars are never completely isolated, and the large energy output of the massive stars certainly has a huge impact on their environment.  These stars interact with their surroundings through their intense ionizing radiation, their powerful stellar winds, and eventually their final supernova explosion. The winds participate in the dissemination of the chemical elements built in the star, but they also transfer large amounts of mechanical momentum  to the InterStellar Medium (ISM). We may note in this context that the total amount of energy released through the wind  during the entire lifetime of a massive star is comparable to the  energy of one supernova explosion: the wind contribution can thus not be neglected. By sweeping up the surrounding gas, the winds are therefore able to shape the ISM into bubbles. These bubbles have sizes of about ten lightyears  when they are blown by single stars, but can reach thousands of lightyears if several of these stars act collectively (they are then called  superbubbles and supergiant shells). This input of mechanical energy into the surroundings can be enough to induce the formation of new generations of stars. \\

One of the first attempts to model these peculiar structures was presented by
Weaver et al.\ (1977, and references therein). This simple model has been 
so often used that it is generally considered as the `standard' one.
In this scheme, a typical bubble consists of four regions, which are, starting from the star outwards, (1) a region where the wind freely 
expands, (2) a zone containing the shocked wind, (3) a region with the 
shocked ISM, and (4) the undisturbed ambient ISM (see Fig.\ \ref{typbub}).\\

The evolution of a typical bubble proceeds mainly through three stages 
(see e.g.\ Lamers \& Cassinelli 1999). At first, the wind is not stopped 
by the ISM, 
and the bubble expands so fast that the radiative cooling does not have 
enough time to affect its evolution: this is called the `adiabatic 
phase'. As the amount of swept-up matter increases, the shock velocity 
begins to decrease, and the cooling 
of the shell of shocked ISM finally becomes significant: the gas compresses 
into a very thin and dense shell. During that phase, the shocked 
wind becomes on the contrary hotter and hotter, and has not enough 
time to cool. The pressure of this hot, shocked wind is so high that 
it now drives the expansion of the shell. 
This marks the beginning of the second phase, during which the bubble 
is generally called an `energy-driven' one. Most observed bubbles are 
in this phase. 
Finally, the hot wind also cools, and in turn collapses into a 
dense shell. At this point, the wind impacts directly on the shell, 
and directly transfers momentum to it:
the bubble is then called a `momentum-conserving' one.\\

  \begin{figure}
  \centering
  \caption{{\it Top:} The structure of a typical wind-blown bubble in the 
energy-driven case (see e.g.\ Weaver et al.\ 1977). $Bottom:$ NGC6888, a bubble blown by a Wolf-Rayet star. \copyright\ SDSU/MLO/Y. Chu et al.}
  \label{typbub}
  \end{figure}

During the second phase, the bubbles are detectable through (nearly) 
the whole electromagnetic spectrum. The stellar wind  flows first
at thousands of \kms, but it nearly stops at the reverse shock that 
expands with velocities of only a few tens \kms. Its entire kinetic energy 
is then converted into thermal energy, heating up the shocked
wind to millions of degrees. The interior of the bubble, filled with 
hot gas, will thus be detectable in X-rays. On the other hand, the 
thin shell of shocked ISM has cooled but is nevertheless typically at 
a temperature of $\sim10^4$~K, since it is still ionized by the stellar 
radiation: it will thus be detectable through the usual nebular emission
lines (\ha, [O\,{\sc{iii}}], ...). 
The interface layer between these two regions is detected through 
absorptions in the UV range, but a gap between the hot interior and 
the dense shell can also reveal the presence of this layer
(Chu et al.\ 2003).\\

The Weaver et al.\ (1977) model attempts to describe the evolution of 
such a bubble. It relies on a few hypotheses: the ISM is 
initially at rest, and it has a uniform density $n_0$; the source 
of the constant wind is point-like and the pressure inside the bubble 
is uniform. 
To find the evolution of an energy-driven bubble, one then simply needs
to apply the conservation equations:\\
\begin{eqnarray*}
& M_{shell} = 4/3\, \pi\, R^3\, n_0\, \mu & {\rm Conservation\ of\ mass} \\
& \frac{d( M_{shell}\,{\rm V}_{exp} ) }{dt} = 4\, \pi \, R^2 \, P & {\rm Conservation\ of\ momentum} \\
& \frac{dE}{dt} = L_w - P \frac{d(4/3\, \pi\, R^3)}{dt} & {\rm Conservation\ of\ energy}\\
&{\rm where\,\,} E = 4/3\, \pi\, R^3\, P/ (\gamma-1)&
\end{eqnarray*}
where $R$, V$_{exp}$ and $M_{shell}$ are the radius, expansion 
velocity and mass of the thin shell of shocked ISM; $n_0$ is the 
ambient number density (in cm$^{-3}$); $\mu$ is the mean molecular weight
in the \hii\ region\footnote{We have used here the convention adopted by Weaver et al. (1977), although in the litterature, the mean molecular weight is often noted $\mu$m$_{\rm H}$. }; $P$ is the pressure inside the bubble; 
$E$ is the internal energy and $\gamma$ is the adiabatic 
index of the hot gas (taken to be 5/3); and $L_w=0.5\, \dot M \, {\rm v}_{\infty}^2$ 
is the mechanical luminosity of the stellar wind.\\

The solution to these three equations is\\
\begin{eqnarray*}
R&=& \left( \frac{125}{154\, \pi} \right)^{1/5} \left( \frac{L_w}{n_0\, \mu} \right)^{1/5} t^{3/5}\\
{\rm V}_{exp} & =& \frac{dR}{dt} = 0.6 R/t\\
P & =& \frac{7}{(3850\, \pi)^{2/5}} L_w^{2/5}\, (n_0\, \mu)^{3/5} t^{-4/5}
\end{eqnarray*}
The temperature and density inside the bubble can further be 
derived if the heat conduction from the shocked wind into the 
shocked ISM is considered (Weaver et al.\ 1977). Once the 
temperature distribution is known, the X-ray luminosity of the 
hot shocked wind can also be estimated (Chu et al.\ 1995). \\

However, when confronted to observations, this model generally experiences
several difficulties. The wind luminosity has to be decreased by an order 
of magnitude to account for the observed shell radius, shell expansion 
and ISM density. In addition, the predicted X-ray luminosity is generally
not compatible with the observations.\\

More specifically, Oey (1996) has performed numerical simulations of 
superbubbles, taking into account the evolution of their stellar content,
including the supernova explosions, and has compared these models to the 
observational properties of several superbubbles. She found that these 
structures could be divided in two classes, one with very discrepant 
expansion velocities, and one whose velocity is more compatible 
with the model. However, even these latter structures require a 
drastic reduction of the input mechanical luminosity of the stars 
to reach a perfect agreement with the simulations. 
Dunne et al.\ (2001) also 
showed that superbubbles are generally brighter in 
the X-ray domain than expected from their stellar content. 
Mass loading and/or off-center supernovae are often thought to be
responsible for these discrepancies. \\

The observed properties of single-star bubbles, like those detected 
around WR stars, have also been investigated and compared to the theoretical 
expectations. Garc\`{\i}a-Segura (1994) has extended the work of
Weaver et al.\ (1977) to this specific case. He has taken into 
account the mass-loss evolution of the massive star prior to 
the WR stage, but his simulations agree more qualitatively than 
quantitatively with the observations. Again, the wind luminosity 
has to be decreased in order to match the observed radius, velocity and X-ray 
luminosity of the bubbles. \\

Superbubbles and WR bubbles are rather complex objects, in which
a lot of poorly known factors (e.g.\ the exact ISM density distribution,
the exact mass-loss history of the star) could influence the shape and 
evolution of the bubble. To get a realistic comparison between the 
so-called `standard' theory and the observations, one should rather 
consider the most simple objects: bubbles blown by a single, main 
sequence massive star interacting directly with the ISM. Such bubbles
are called `interstellar bubbles', but only a few of them are known. Naz\'e et al. (2001b, 2002) have discovered several such structures in N11B, N180B, and N44, and their properties agree better with theoretical expectations than in the case of WR bubbles and superbubbles, but the agreement is still far from perfect.\\

\section{Spectroscopy of Single Massive Stars}
Although many wavelength ranges (e.g. infrared and ultraviolet) provide important spectroscopic information, we will focus in this section on the results obtained in the visible and X-ray domains. 

\subsection{Visible Domain}
Apart from classifying the star, visible spectra are also used as input for modelling and variability studies.\\

Modelling spectra constitutes a crucial step in astrophysics since it is the only way to derive intrinsic stellar properties like temperature, mass-loss rate, gravity, chemical composition, rotation,... Even distances could be estimated thanks to modelling: once the wind-momentum luminosity relation (see above) is calibrated, the determination of the radius, mass-loss rate and terminal velocity through the modelling of the observed stellar spectrum leads to an estimate of the intrinsic luminosity of the star, hence its distance. Although modelling stellar spectra is not a trivial task, it has nevertheless greatly improved in recent years. The first models were indeed rather simple: they assumed plane-parallel and static atmospheres composed only of hydrogen and helium. This led to the first determinations of the temperature scales of massive stars. Very soon, a problem arose: the mass predicted by these models was systematically smaller than that predicted by evolutionary models on the basis of the position of the star in the HR diagram (Herrero 2005 and references therein). This so-called mass discrepancy was mostly solved by new, improved models that include:\\
\begin{itemize}
\item {\it non-LTE (Local Thermodynamic Equilibrium) effects}. Since massive stars possess a very intense radiation field, the radiative phenomena largely dominate over the collisional effects.
\item {\it a spherical dynamic atmosphere}. Plane-parallel approximation is valid only if the height of the atmosphere is small compared to the stellar radius. This is not the case for massive stars, since the optical depth of the wind can be significant out to several tens of stellar radii. In addition, the windy atmosphere of massive stars is expanding, and a photon emitted at one point can be absorbed much farther by another line thanks to the Doppler effect.
\item {\it the line-blanketing}. Metals, although rare, play an important role in the ionisation of the wind since they are very efficient at absorbing photons. This is especially true in the UV, where numerous lines are present: different atoms/ions can thus absorb radiation in the same frequency range. The inclusion of metals in the modelling thus results in a blocking of the UV flux, that leads (1) to a heating of the inner atmosphere (backwarming) and (2) to the cooling of the outer atmosphere (where the ionisation, determined by the now reduced UV flux, is decreased).
\end{itemize}
Such models have led to a new parameter scale for massive stars (see Martins et al. 2005, parameters reproduced in Table \ref{martins}). Due to the backwarming effect of the inner atmosphere where the absorption lines form, the same ionisation (i.e. helium ratios) can be reached with a lower effective temperature, and this leads to a reduction of the temperature scale by up to 8000K. On the contrary, for Wolf-Rayet stars, the emission lines are formed in the outer parts of the atmosphere, where the inclusion of the line blanketing results in a reduction of the ionisation: higher effective temperatures are thus needed to fit the spectra. However, we may mention that the most recent models are still 1-D and stationary ones, and that the work continues to further improve  the models.\\

\begin{table}
\caption{Theoretical stellar parameters as a function of spectral types and luminosity classes, as determined by Martins et al. (2005). The effective temperature $T_{eff}$, here displayed in kK, is the temperature of a blackbody emitting the same amount of radiation as the star (therefore $L=4\pi R_*^2\sigma T_{eff}^4$); $R_*$ is the stellar radius in \rsol\ and  $M_*$ the stellar mass in \msol.
\label{martins}} 
\small
\begin{center} 
\begin{tabular}{l | r r r | r r r | r r r}
\hline 
Subtype & \multicolumn{3}{|c|}{Dwarf Stars (V)}  & \multicolumn{3}{c|}{Giant Stars (III)}  & \multicolumn{3}{c}{Supergiant Stars (I)}\\
& $T_{eff}$ & $R_*$ & $M_*$ & $T_{eff}$ & $R_*$ & $M_*$ & $T_{eff}$ & $R_*$ & $M_*$ \\
\hline\hline
O3  & 44.6& 13.8& 58.3& 42.9& 16.6& 58.6& 42.6& 18.5& 66.9\\
O4  & 43.4& 12.3& 46.2& 41.5& 15.8& 48.8& 40.7& 18.9& 58.0\\
O5  & 41.5& 11.1& 37.3& 39.5& 15.3& 41.5& 38.5& 19.5& 50.9\\
O5.5& 40.1& 10.6& 34.2& 38.0& 15.1& 38.9& 37.1& 19.9& 48.3\\
O6  & 38.2& 10.2& 31.7& 36.7& 15.0& 36.4& 35.7& 20.3& 45.8\\
O6.5& 36.8&  9.8& 29.0& 35.6& 14.7& 33.7& 34.7& 20.7& 43.1\\
O7  & 35.5&  9.4& 26.5& 34.6& 14.5& 31.2& 33.3& 21.1& 40.9\\
O7.5& 34.4&  8.9& 24.2& 33.5& 14.3& 29.1& 31.9& 21.7& 39.2\\
O8  & 33.4&  8.5& 22.0& 32.6& 14.1& 26.9& 31.0& 22.0& 36.8\\
O8.5& 32.5&  8.1& 19.8& 31.7& 13.9& 24.8& 30.5& 22.2& 33.9\\
O9  & 31.5&  7.7& 18.0& 30.7& 13.7& 23.1& 29.6& 22.6& 32.0\\
O9.5& 30.5&  7.4& 16.5& 30.2& 13.4& 21.0& 28.4& 23.1& 30.4\\
\hline 
\end{tabular} 
\end{center} 
\end{table} 

Another possible analysis of the spectrum of a single star is to investigate its variability. A very efficient tool for this task is the Temporal Variance Spectrum (TVS) that was defined by Fullerton et al. (1996). Consider a dataset of $N$ normalized spectra with the same wavelength sampling and arrange them in a matrix $S$ where $S_{ij}$ is the jth wavelength element of the ith spectrum. To search for variability, the spectra have to be compared to a mean spectrum, and the differences tested statistically. However, the signal-to-noise ratio is different for each spectrum and a weighting is therefore needed in order to perform meaningful statistical tests. To this aim, we define the weighting factor $w_i$ as $\sigma_0^2/\sigma_{ic}^2$, where $\sigma_{ic}$ is the inverse of the signal-to-noise in the continuum of the ith spectrum, and $\sigma_0^2$ is equal to $\left( \frac{1}{N} \sum_{i=1}^N \sigma_{ic}^{-2} \right)^{-1}$, i.e. $\sigma_0$ is the inverse of the rms signal-to-noise in the continuum of the dataset. The weighting factor therefore enables to reduce the importance of low quality (i.e. low signal-to-noise) spectra. It is also important to take into account the wavelength-to-wavelength variations of the noise: indeed, the signal-to-noise ratio is higher in emission lines, but the intrinsic variations are also higher and larger deviations are thus normally expected for these lines. The observed deviations have thus to be ``normalized'' by a correcting factor reflecting the expected noise. If the exposure time is sufficient, the instrumental noise is negligible compared to the photon noise and a good correction factor $\alpha_{ij}=\sigma_{ij}^2/\sigma_{ic}^2$ would be simply $S_{ij}$, the spectrum itself: $\alpha_{ij}$ is then $<1$ if there is an absorption line (since a lower signal implies a lower noise) and $>1$ in the case of an emission line above the continuum. Taking these correcting factors into account, the TVS at a given wavelength is then defined as:\\
\hspace*{2cm} $TVS_j=\frac{1}{N-1} \sum_{i=1}^N \frac{w_i}{\alpha_{ij}} (S_{ij}-\overline{S_j})^2$ \\
\hspace*{4.9cm} with the mean spectrum $\overline{S_j}=\frac{1}{N} \sum_{i=1}^N w_i S_{ij}$\\
This TVS follows a distribution $\sigma_0^2 \chi^2_{N-1}$ ($\chi^2$ being here the reduced chi-squared), and a statistical test can therefore be performed easily: the deviations are generally considered significant if they exceed the 99\% level. An example of such a TVS is shown in Fig.\ \ref{tvs}.\\

  \begin{figure}
  \centering
  \caption{Mean visible spectrum of HD\,191612 (top) and Time Variance Spectrum (TVS, bottom). (From Naz\'e et al.\ 2005)
\label{tvs}}
  \end{figure}

  \begin{figure}
  \centering
  \caption{A corotating interaction region, with a spiral form (dashed line), can modify the spectrum of massive stars. A component will first appear in the line of sight at intermediate velocities (case A, solid line). As the star revolves, only parts of the spiral can be seen at low and high velocities (case B, solid lines) and the observed absorption component therefore seems to migrate towards long and short wavelengths, the relative importance of the blueshifted vs.\ redshifted component depending on the physical parameters of the region and the star (e.g. rotation rate, change in the mass-loss rate and/or rotation).  \label{spiral}}
  \end{figure}

  \begin{figure}
  \centering
  \caption{ Schematic view of the $\theta^1$\,Ori\,C system.
\label{thetaori}}
  \end{figure}

  \begin{figure}
  \centering
  \caption{Variations of the EW of the \ha\ line (top), of the \hei\,\l\,4471 and \heii\,\l\,4542 lines (middle) and Hipparcos photometry of \hd191612 (bottom). (From Naz\'e et al.\ 2005) \label{191612}}
  \end{figure}  

Once a variability is detected and if the spectra are sufficiently numerous, the dataset can be searched for periodicities in the observed variations. This can be done for example with the Generalized Fourier Transform (see Heck et al. 1985 and remarks in Gosset et al. 2001) that extends the Fourier Transform to the case of non-regular sampling.\\

For massive stars, a stochastic variability is generally expected because of the presence of the unstable stellar wind that generates short-lived small-scale structures. In fact, the instability of the wind is intimately linked to the line-driven mechanism. Consider an atom or ion at a distance $r$ of the star and moving at a velocity $v_r$. It absorbs photons at a frequency $\nu_0$ in its reference frame, i.e. a frequency $\nu_0 (1+v_r/c)$ in the reference frame of the star (cf. above). The velocity can be slightly perturbed and become $v+\delta v$. If $\delta v >0$, the atom/ion will absorb photons of higher frequency, of which plenty are available, and it will therefore accelerate even more: the perturbation is thus amplified. On the contrary, if $\delta v <0$, the atom/ion can only absorb photons of lower frequency but these have already been absorbed by the slower material closer to the star and are no more available: the particle will therefore decelerate even more with respect to the unperturbed $v_r$ velocity law, and the perturbation is again amplified. Hence, any slight perturbation of the velocity is doomed to be amplified, provoking the formation of small-scale structures (so-called clumps) in the wind. These structures are thought to produce a stochastic variability of the spectrum.\\

Sometimes, however, variations appear to be regular, with periods ranging from a few hours to several years. The most common sources of variability include:\\ 
\begin{itemize}
\item {\it Pulsations}. It is well known that the spectra of Cepheids change as these stars pulsate radially. Non-radial pulsations of lower intensity can also modify the line profiles and magnitudes of stars. Generally, asteroseismology has focused on low-mass stars, but a few massive objects (e.g. $\zeta$ Oph, HD\,152219, HD\,93521) apparently also display pulsations.  
\item {\it Structures in the Wind}. Structures in the wind on a rather large scale can appear and disappear, modifying the observed spectrum. For example, if the stellar surface harbors a cold (resp. hot) point, the mass-loss rate above that point will be decreased (resp. increased) and the velocity of the gas will become larger (resp. smaller) because of the reduced (resp. increased) absorption: this modified wind will soon collide with the ``normal" surrounding gas. Due to the stellar rotation, spiral structures might then appear, and give rise e.g. to Discrete Absorption Components (DACs, see Fig.\ \ref{spiral} and Cranmer \& Owocki 1996).
\item {\it Magnetic Fields}. Because of the very broad lines in the spectra of the massive stars, it is very difficult to estimate their magnetic field and so far there are only two cases, the stars $\theta^1$\,Ori\,C and HD\,191612, where it has been measured with certainty. In the $\theta^1$\,Ori\,C system, the wind material is funneled by the magnetic field towards the magnetic equator, creating a dense equatorial region where some emission lines can arise. A recurrent modulation of the spectrum then appears because the magnetic axis is not aligned with the rotation axis (hence the name {\it magnetic oblique rotator} used for $\theta^1$\,Ori\,C): different parts of the ``disk" are therefore seen at different phases of the rotation cycle (see Fig.\ \ref{thetaori} and Stahl et 
al.\ 1996).
\end{itemize}

In this context, the peculiar variability of the Of?p stars needs to be mentioned. The Of?p category was introduced by Nolan Walborn in 1972 to describe two stars, \hd108 and \hd148937, with spectra that were slightly different from those of normal Of supergiants. Notably, they present \ciii\ lines around 4650\AA\ with an intensity comparable to that of the neighbouring \niii\ lines. In addition, their spectra show sharp emission lines and some P Cygni profiles. A third star was soon added to this new class, \hd191612. The observation of \hd108, the best studied member of this class, led to conflicting results in the past, with explanations for the radial velocity variations ranging from binary motion to stochastic wind instabilities. Using a 15yr monitoring campaign of the star, Naz\'e et al. (2001a) discovered that the star actually underwent long-term line profile variations: the Balmer lines and the \hei\ lines passed from emission or P Cygni profiles to absorptions while other emission and absorption lines were unchanged, like \heii\,\l\,4542. These variations appear recurrent with a timescale of approximately 50-60 years. A few years later, Walborn et al. (2003, 2004) reported a very similar phenomenon in the spectrum of another Of?p star, \hd191612 (see Fig.\ \ref{191612}), but the timescale appears much shorter, about $\sim$540~days. The same timescale was subsequently detected in Hipparcos photometry (see Naz\'e et al. 2005 and references therein). Investigations to determine the exact nature of these peculiar stars are still ongoing. 

\subsection{X-rays}

Because of technical difficulties, X-ray astronomy was born quite recently. Indeed, rockets or satellites are needed to overcome the absorbing effect of our atmosphere - doing X-ray astronomy was thus not possible before World War II. In addition, the efficiency of X-ray detectors and telescopes has improved very slowly, and that explains why the first generation of ``great observatories" (i.e. \ch\ and \xmm) was only launched in 1999.\\

  \begin{figure}
  \centering
  \caption{ X-ray spectrum of 9Sgr, obtained with \xmm\ using a CCD (top) or a grating (bottom). In the low resolution spectrum, the lines seen at high resolution are blended, forming a bell-shape pseudo-continuum. (from Rauw et al. 2002) \label{specx}}
  \end{figure}  

Spectroscopy in X-rays can be performed in 3 different ways:\\
\begin{itemize}
\item {\it CCDs}. CCDs for X-rays can provide much more than a simple image. Due to the low luminosities of astronomical objects in the X-ray range, the X-ray photons can in fact be recorded one at a time. The electron shower generated at the arrival of the photon will therefore be recorded precisely in position and also in intensity. As the number of electrons is directly proportional to the energy of the incident photon, CCDs provide a cheap and simple way to do spectroscopy, although only at a rather low-resolution ($R=dE/E\sim 10-50$).
\item {\it Gratings}. Gratings (either in transmission or reflection) can be used in the X-rays with only little modification compared to the visible domain. Such instruments provide a higher resolution than CCDs: $R\sim 200-2000$.
\item {\it Bolometers}. As for CCDs, X-rays are detected once at a time in bolometers, where the photon energy is converted into thermal energy of the electrons. Since a higher frequency photon will lead to a larger temperature increase of the bolometer, spectroscopy is a direct by-product of the use of bolometers. They provide high-resolution spectra ($R\sim 1000$), but they need to be cooled to 0.1K (because an X-ray photon will provoke a $\Delta T$ of only a few 0.001K!). The Japanese observatories Astro-E and Astro-E2 should have used the first bolometers for X-ray astronomy, but the former exploded after launch and the latter lost all its liquid helium (necessary to cool the bolometer) shortly after orbit insertion.
\end{itemize}
An example of spectra obtained with the first and second methods is shown in Fig.\ \ref{specx}.\\

As far as massive stars are concerned, X-ray astronomy really began 25 years ago. At that time, the $Einstein$ observatory had just been  launched and NASA was trying to calibrate it by observing well-known sources. The observation of one of these sources, Cyg X-3, revealed four nearby spots  (see Fig.\ \ref{cygx3}). At first thought to be due to an instrumental effect, the spots were soon found to mark the discovery of X-ray emission from 4 massive stars belonging to the Cyg OB2 star cluster. Indeed, $Einstein$ and its followers showed that X-ray emission is very common among massive stars. \\

  \begin{figure}
  \centering
  \caption{The X-ray emission from massive stars was discovered serendipitously in December 1978, when $Einstein$ observed the bright Cyg X-3 for calibration purposes. The four ``spots" above Cyg X-3 correspond to the massive stars Cyg OB2 \#5, 8, 9 and 12. \copyright\ $Einstein$}
  \label{cygx3}
  \end{figure}

However, the exact origin of that emission is still under debate. Some authors had first proposed that a corona at the base of the wind, analogous to what exists in low-mass stars, could be responsible for the high-energy emission of massive stars.  However, several observational objections against such models were raised (see Owocki \& Cohen 1999 and  references therein): absence of a strong attenuation by the stellar wind (this suggests  that the source of the X-ray emission lies significantly above the  photosphere, at several stellar radii), too low an X-ray output,  inconsistencies between UV and X-ray predictions compared to  observations,... As an alternative, a scenario based on the instability of the line-driven mechanism  has been proposed. In fact, an unstable line-driven wind does not flow at the same velocity everywhere and shocks between the different parts are expected, causing the formation of dense shells which will be distributed throughout the whole wind. At first, the forward shocks between  a fast wind and the ambient slow (``shadowed'') material were  considered as the probable cause of X-ray emission, but subsequent  hydrodynamical simulations rather showed the presence of strong  reverse shocks which decelerate the fast, low density material. However, the resulting X-ray emission  from such material after it crossed the reverse shock is very low,  and can probably not explain the level of X-ray emission observed  amongst O stars. More recent simulations by Feldmeier et al.\ (1997) have shown that mutual collisions of dense shells of gas compressed in the shocks would lead to substantial X-ray luminosities,  comparable to the observed ones. Such models also predict significant  short-term variations of the X-ray flux but since this is not  observed, it was concluded that the winds are most probably fragmented (or clumpy),  so that individual X-ray fluctuations are smoothed out over the whole  emitting volume, leading to a rather constant X-ray output (Feldmeier et al.\ 1997).  In addition, a supplementary X-ray emission may result from  other mechanisms, which are not necessarily present in every massive star.  For example, an accreting compact object will generally emit  a wealth of X-rays and can have a drastic impact on the stellar wind  structure (e.g.\ Kallman \& Mc Cray 1982). In binary systems  containing two hot stars, a colliding-wind phenomenon (see next section) and/or inverse Compton scattering  by relativistic electrons accelerated by the shocks (Chen \& White 1991) will also lead to additional X-ray emission. Finally, the wind from both hemispheres, deviated by a magnetic field, can collide in the equatorial regions, and provide  another substantial source of X-ray emission (Babel \& Montmerle 1997). \\

On the observational side, it was soon found that the X-ray luminosity scales with the bolometric luminosity of massive stars. Although quite dispersed in the past, the data now show the relation to be rather tight: $\log (L_X\, in \,0.5-10.\,keV) =\log(L_{BOL})-6.91\pm0.15$ (Sana et al. 2006b). On the other hand, low-resolution spectroscopy has unveiled the fundamental characteristics of the X-ray emission. First, it is not of the blackbody-type: as the heated gas is optically transparent at these wavelengths, the observed emission corresponds to the superposition of discrete emission lines. This emission can thus be fitted by optically thin thermal plasma ``mekal" or ``Raymond-Smith" models. Moreover, the temperature of the emitting gas is found to be about 0.5keV, or 6MK, and the wind absorption is generally low, except for Wolf-Rayet stars.\\

The high-resolution spectra enable to go further into the analysis of the X-ray emission. Two main types of studies can be undertaken. The first one is related to the $fir$ triplets. These lines are seen in helium-like ions and correspond to transitions from the $n=2$ level to the $n=1$ ground level. The $f$ line, or forbidden line, arises from the transition 1s 2s ($^3S_1$) $\rightarrow$ 1s$^2$ ($^1S_0$) whereas the $r$ (for resonance) line is linked to the transition 1s 2p ($^1P_1$) $\rightarrow$ 1s$^2$ ($^1S_0$) and the $i$ (for intercombination) line to the transition 1s 2p ($^3P_1$) $\rightarrow$ 1s$^2$ ($^1S_0$). In stellar coronae, the ratio of the $f$ and $i$ lines mainly depends of the electron density. However, in the hot plasma surrounding massive stars, the UV radiation plays an important role by coupling the upper levels of these two lines thereby reducing the $f$ line in favor of the $i$ line. Therefore, the $\Re = f/i$ ratio is a diagnosis of the dilution factor $0.5 \left( 1- \sqrt{1-\left( \frac{R_*}{r}\right) ^2} \right)$ of the UV radiation, i.e. a diagnosis of the distance $r$ from the stellar surface where the X-rays are emitted (see Fig.\ \ref{fir}). For example, the X-rays from $\zeta$\,Pup and 9\,Sgr apparently form at a few stellar radii, as predicted by the standard model\footnote{If the X-ray emitting material is distributed throughout the wind, the lines should form near the radius where $\tau=1$ at the specific energy of the line. This is the case for 9 Sgr (see right part of Fig.\ \ref{fir}).}, whereas the $f/i$ ratio observed for $\theta^1$\,Ori\,C suggests that the X-ray emission arises very close to the photosphere (Rauw 2005b). Note that the $G=\frac{f+i}{r}$ ratio is sensitive to the temperature of the gas.\\

  \begin{figure}
  \centering
  \caption{Analysis of the $fir$ lines of 9 Sgr. {\it Top:} Theoretical $f/i$ ratio for different helium-like ions with the 90\% confidence interval of the observed ratio for Ne\,{\sc ix} (thick solid line). {\it Bottom:} Radius of optical depth unity as a function of wavelength: the observed ratio is compatible with the radius of unity optical depth. (from Rauw et al. 2002) \label{fir}}
  \end{figure}  

In addition, high-resolution spectroscopy also enables to investigate the detailed morphology of the X-ray lines, leading to additional physical information. If the wind were optically thin without any absorption, the lines would appear flat-topped (see right part of Fig.\ \ref{morph}). However, the wind absorbs part of the X-rays that it has emitted itself. This absorption will be larger if there is more material between the emission region and the observer, as is the case for X-rays emitted in the receding part of the wind. The absorption will thus affect more the red part of the line: the observed line will therefore appear blueshifted (see right part of Fig.\ \ref{morph}). In addition, the wind expands with a large velocity, so that the lines should appear rather broad. Such broad, asymmetric lines are observed for $\zeta$ Pup and 9 Sgr. However, $\zeta$ Ori displays broad but symmetric lines (Fig. \ref{morph}, Rauw 2005b). To explain this difference, Oskinova et al. (2005) have proposed to consider a porous wind consisting of optically-thick clumps: although each clump can absorb efficiently the radiation, the X-rays can still escape freely by passing between them. Therefore, the absorption does not depend on the opacity of the wind (clumps), but on the spatial distribution of the clumps. If the clumps are tightly packed, the wind is nearly homogeneous and the lines will be skewed. On the other hand, if the clumps are rare and distant, the lines will be symmetric. Finally, $\delta$ Ori presents narrow symmetric lines that could be due to a colliding wind phenomenon (see below) while the lines of $\theta^1$\,Ori\,C can be easily explained by the confined wind model (Rauw 2005b).

  \begin{figure}
  \centering
  \caption{X-rays lines at high resolution. {\it Top:} $\zeta$ Pup (from Cassinelli et al. 2001)  {\it Bottom:} $\zeta$ Ori (from Waldron \& Cassinelli 2001) \label{morph}}
  \end{figure}

\section{Spectroscopy of Massive Binaries}
\subsection{Colliding Winds : Introduction}
As we have seen before, massive stars blow dense and powerful stellar winds. If such stars belong to a binary system, a collision between the two winds is unavoidable. Since the winds are supersonic, the shock between them is a strong one and the gas will become very hot and dense after the collision. This phenomenon was predicted a few decades ago, but it was only considered seriously since recent years, when the observational evidence began to accumulate. Some theoretical considerations will first be presented, before reviewing the observational data. \\

The temperature of the gas after the shock can be evaluated by $T=\frac{3\mu v^2_{\perp}}{16 k}$ (with $\mu$ the mean weight of wind particles and $v_{\perp}$ the component of the pre-shock wind velocity perpendicular to the shock surface). For solar abundances, this equation can be written as $T({\rm K})\sim 1.4\,10^7 v^2_{\perp,8}$ or $kT ({\rm keV})\sim 1.2 v^2_{\perp,8}$, if $v_{\perp,8}$ is the velocity $v_{\perp}$ expressed in thousands of \kms. For typical winds ($v_{\perp}\sim2000$\kms), the temperature can reach 60MK. To understand the physical properties of the post-shock gas, we can use the ratio between the characteristic timescale of radiative cooling and the time to escape the shock zone (Stevens et al. 1992): $\chi=\frac{T_{cool}}{T_{esc}}\sim\frac{v^4_{\perp,8} x_7}{\dot M_{-7}}$ where $x_7$ is the separation between the considered star and the stagnation point (the intersection of the contact surface with the axis joining the two stars' centers), expressed in $10^7$km and the mass-loss rate $\dot M$ should be given in $10^{-7}$\msol yr$^{-1}$. For values of $\chi$ close to or larger than one, the cooling does not have the time to play a role and the shock can be considered as adiabatic: the gas remains at a high temperature. In this case, no optical emission line should form in the collision zone. On the contrary, when $\chi<<1$, the cooling is very efficient and the shock radiates a lot (i.e. emission lines over a broad range of wavelength, including the optical domain, will now be generated in the shock zone). In this case, the shock zone is compressed and subject to many instabilities (see Fig.\ \ref{simulations} and Stevens et al. 1992).\\

  \begin{figure}
  \centering
  \caption{Hydrodynamical simulations for \hd152248 (left, radiative case) and \hd93403 (right, adiabatic case). (H. Sana, private communication) \label{simulations}}
  \end{figure}

The geometry of the collision zone can also be derived rather easily since the contact surface of the two winds corresponds to the equilibrium between the two wind ram pressures (Stevens et al. 1992 and Fig.\ \ref{cwregion}):\\ 
\hspace*{2cm} $ \rho_{1}\,v_{\perp,1}^2 = \rho_{2}\,v_{\perp,2}^2 $\\
\hspace*{2cm} or $ \rho_{1}\,v_{1}^2\cos^2{\phi_{1}} = \rho_{2}\,v_{2}^2\cos^2{\phi_{2}} $ \\
Since the continuity equation implies ${\dot M} = 4\,\pi\,r^2\,v\,\rho$, the above relation can be re-written as:\\
\hspace*{2cm} $\frac{\dot {\rm M}_{1}\,v_{1}}{4\,\pi\,r_{1}^{2}}\,\cos^2{\phi_1} = \frac{\dot {\rm M}_{2}\,v_{2}}{4\,\pi\,r_{2}^{2}}\,\cos^2{\phi_2}$\\

As we have seen before, the velocity in the wind follows a $\beta$-law, i.e. $v(r)= v_{\infty}(1-\frac{R}{r})^{\beta}$ and we can therefore write: \\
\hspace*{2cm} $ \cos{\phi_2} = {\mathcal R}\,\frac{r_2}{r_1}\,\frac{(1 - R_1/r_1)^{\beta_1/2}}{(1 - R_2/r_2)^{\beta_2/2}}\,\cos{\phi_1}$ \\
if we define the on-axis momentum ratio as
${\mathcal R} = \left( \frac{\dot M_{1}\, v_{\infty, 1}}{\dot M_{2}\, v_{\infty, 2}} \right)^{1/2}$.\\
The equation is indeed equivalent to:\\
\hspace*{2cm} $\cos{\phi_2} = \lambda\,\cos{\phi_1}\,\frac{r_2}{r_1}$\\
where $\lambda = {\mathcal R}\,(1 - R_1/r_1)^{\beta_1/2}/(1 - R_2/r_2)^{\beta_2/2}$.\\

  \begin{figure}
  \centering
  \caption{Geometry of the colliding wind region. (G. Rauw, private communication) \label{cwregion}}
  \end{figure}

  \begin{figure}
  \centering
  \caption{Contact surfaces for different momentum ratios.  (From Sana \& Rauw 2003)
\label{cwregionbis}}
  \end{figure}

In addition, we know that $\pi/2 - \phi_1 = \beta - \theta_1$ and $\pi/2 - \phi_2 = \theta_2 - \beta$ (see Fig.\ \ref{cwregion}), and therefore\\
\hspace*{2cm} $\cos{\phi_1} =  \sin{\beta}\,\cos{\theta_1} - \cos{\beta}\,\sin{\theta_1}$ \\
\hspace*{2cm} $\cos{\phi_2} = \cos{\beta}\,\sin{\theta_2} - \sin{\beta}\,\cos{\theta_2}$\\

The equation then becomes \\
\hspace*{2cm} $ \frac{r_1}{r_2} = \lambda\,\frac{\tan{\beta}\,\cos{\theta_1} - \sin{\theta_1}}{\sin{\theta_2} - \tan{\beta}\,\cos{\theta_2}}$\\
And finally, the equation of the contact surface is \\
\hspace*{2cm} $\frac{dz}{dx} = \tan{\beta} = \frac{(\lambda\,r_2^{2} + r_1^2)\,z}{\lambda\,r_2^2\,x + r_1^2\,(x - d)}$\\
with $ r_1 = \sqrt{x^2 + z^2}, r_2 = \sqrt{(x-d)^2 + z^2}$\\

If the winds have reached the terminal velocity before they collide, $\lambda$ simplifies to ${\mathcal R}$. In this specific case, the on-axis momentum ratio ${\mathcal R}$ can be easily physically interpreted. First, it enables to find the stagnation point. In fact, in this case, the two angles $\phi_1$ and $\phi_2$ are zero and ${\mathcal R}$ directly gives the ratio of the distances between the stagnation point and the stars. Therefore, the stagnation is closer to the star with the weaker wind (see Fig.\ \ref{cwregionbis}). In addition, ${\mathcal R}$ also gives the form of the shock because the half opening angle of the shock cone (in $^{\circ}$) was found empirically to be $\sim 120\left( 1-\frac{{\mathcal R}^{-4/5}}{4} \right){\mathcal R}^{-2/3}$. The shock thus wraps around the star with the weaker wind (see Fig.\ \ref{cwregionbis}).\\

In the above discussion, several effects have been neglected. For example, in some tight binary systems, there will be a deflection of the shock zone because of the orbital motion. Moreover, if the system is eccentric, the colliding wind (CW) phenomenon could appear only near periastron, i.e. when the stars are closer to each other. In addition, since the winds are driven by the scattering of UV photons, the wind of one star is affected by the presence of the radiation field of the other star. This effect is called `radiative inhibition': the radiative pressure from a companion being able to slow the radiation-driven winds, a weaker CW shock will result. Finally, the asymmetry or clumpiness of the winds has also been neglected and only starts to be taken into account in some hydrodynamical simulations.\\

\subsection{Signature in the Visible Range}
A binary can be easily detected with spectroscopy, except if the system is seen face-on, since the lines regularly shift from the blue to the red side of the spectrum, in harmony with the orbital motion. If the moving lines of only one star are seen, one talks about an SB1 (spectroscopic binary with one component detectable); if the lines of both stars are observed, the system is called a SB2. Normally, all lines detected in the composite spectrum of the system belong to one star or the other but, in the case of a radiative CW, additional emission lines appear. Since these lines are not formed in or close to the photosphere of the stars, they do not follow the orbital motion of the system.\\

  \begin{figure}
  \centering
  \caption{Definition of the axes for tomographic analysis (see text).   \label{tomodef}}
  \end{figure}

Doppler tomography can help to better determine the properties of these peculiar emissions but to apply that technique, a good spectroscopic coverage of the orbital cycle is crucial. Two versions of the Doppler tomography are available: the simple S-wave analysis and the more sophisticated Doppler mapping (Rauw 2005a). Both request the definition of specific axes $(x,y)$: $x$ is the binary axis, from the primary to the secondary star; $y$ is perpendicular to $x$, and in the direction of the motion of the primary star (see Fig.\ \ref{tomodef}). These axes are not fixed but rather rotate with the orbital motion. If an emission component (either a discrete one or the position of the peak of a broad emission line) has fixed velocities in this reference frame, it will appear with a velocity $v(\phi) = V(\phi) \sin i = -V_x \cos(2\pi \phi) \sin i+V_y \sin(2\pi \phi) \sin i$ or $-v_x \cos(2\pi \phi) +v_y \sin(2\pi \phi)$ throughout the orbital cycle ($\phi$ being the orbital phase). The above combination of a cosine and a sine generates a radial velocity curve that displays an S-shape in the dynamical spectrum, hence the name `S-wave analysis'. In practice, one thus tries to fit an S-shape function to the observed radial velocities of the chosen emission component. Then a Doppler map reporting the fitted $(v_x,v_y)$ is created. This map generally also shows the star's position in the velocity space, i.e. $(0,-K_1)$ for the primary and $(0,K_2)$ for the secondary, considering that the system is not eccentric (the stars never get closer to or farther away from each other). The velocity equivalent of the Roche lobes can also be displayed, as is the case in Fig.\ \ref{hd149404}.\\

To analyze broad emissions, a tool first developed for medical imaging must be used: the Radon transform. It is defined as $g(s,\theta) = \int_{-\infty}^{+\infty} f(j,k) dt$ where $(s,t)$ and $(j,k)$ are two orthogonal reference frames with an angle $\theta$ between $j$ and $s$, i.e. $j=s \cos (\theta)-t \sin(\theta)$ and $k=s \sin (\theta)+t \cos(\theta)$. In astronomical spectroscopy, $g(s,\theta)$ is actually $I(v, \phi)$, i.e. the intensity of a spectrum at a given velocity and a given phase, which results from the integration of the emissivity function along the line-of-sight through the $(v_x,v_y)$ space. As a consequence, one does not use the Radon transform to get Doppler maps, but the {\it inverse} Radon transform. Practically, the spectra should first be filtered to suppress the high-frequency noise which would degrade the Point Spread Function of the resulting map. Next, for each observed phase, one must find the intensity in the spectrum at a velocity corresponding to each pair $(v_x,v_y)$. Finally, the resulting intensities along the orbital cycle are added, using a weighting to take into account the different phase intervals covered by each observation (Rauw 2005a). If the emission corresponds preferentially to a specific region in the $(v_x,v_y)$ space, a peak will appear at that position of the Doppler map; on the contrary, the signal at other velocities appears with random intensities and will therefore cancel out (see Fig.\ \ref{lss3074}).\\

While Doppler tomography can be very useful, it must be reminded that the Doppler maps display the velocity field. They are NOT `usual', spatial maps, and should thus not be interpreted as such: components close in the velocity space can actually be very distant in position. Another example is the Doppler map of an accretion disk: the higher velocities are reached closer to the star, therefore, the inner (resp. outer) part of the disk will appear on the outside (resp. inside) in the velocity space! In addition, the tomography should be applied with care to eclipsing binary systems (the relative importance of the components would then be biased). The same holds for the systems where emission arises outside the orbital plane (which is unfortunately the case for winds of massive stars).\\

  \begin{figure}
  \centering
  \caption{ Doppler map showing the results of the S-wave analysis for \hd149404 (O7.5I+ON9.7I). The crosses indicate the velocity of the center of mass of the binary components while the dashed line is the equivalent of the Roche lobe in the velocity space. The different symbols stand for different emission lines: filled circles = \ha; open circles = H$\beta $; filled triangles = \nii\l\l5932,5942; open triangles = \niii\l\l4634-41; filled squares = \s\l\l4486,4504; open squares = \ciii\l5696 and stars = \heii\l4686. (From Rauw et al. 2001) \label{hd149404}}
  \end{figure}

  \begin{figure}
  \centering
  \caption{Doppler tomography of the \heii\l4686 line of LSS3074 (O4f$^+$+O6). (G. Rauw, private communication)  \label{lss3074}}
  \end{figure}

On the other hand, other signatures of the CW phenomenon can also be detected. For example, in \hd152248 (O7.5III+O7III), Sana et al. (2001) discovered that the strength and width of some emission lines was varying (see Fig.\ \ref{hd152248vis}). In fact, these emission lines were broader and stronger at quadrature (i.e. when the stars have maximum radial velocities) than at conjunction. This can be easily explained by the presence of a planar\footnote{since both stars have very similar stellar and wind properties} CW region just between the two stars. The emission components produced in this dense CW region would be occulted at conjunction phases: the equivalent width of the lines should be smaller then. In addition, since the shock is almost perpendicular to the axis of the system, the radial velocities of the particles escaping from the wind interaction region show a broader distribution when our line of sight is aligned with the interaction zone (i.e. at quadrature) than at conjunction.\\

  \begin{figure}
  \centering
  \caption{Full widths of the base of the emission component of the \heii\l4686 (left) and H$\alpha$ (right) lines of \hd152248. Different symbols refer to different instruments. Note that the absorption components of the stars have been subtracted from the original line before measuring its width. (From Sana et al. 2001)  \label{hd152248vis}}
  \end{figure}

\subsection{Signature in the X-rays}
In view of the high post-shock temperature, X-ray emission appears as an obvious signature of CW phenomena. Such signatures have indeed been found in many binaries, and they are generally characterized by:\\ 
\begin{itemize}
\item {\it Large X-ray luminosity}. As CW represent an additional phenomena, the X-ray luminosity of binary systems with CW should exceed the simple combination of the two individual X-ray luminosities of the massive stars. Therefore, when comparing X-ray luminosities to bolometric luminosities, CW systems appear above the `classical' $L_X-L_{BOL}$ relation (see Fig.\ \ref{lxlbol}).
\item {\it High temperature}. The X-ray emission from massive stars generally displays a rather low temperature $kT$ (about 0.5keV). However, the emission from CW arises in hotter plasma and should therefore present higher $kT$. Non-thermal emission coming from inverse Compton scattering (see below) could also be observed in colliding wind binaries (CWB).
\item {\it Modulation of the X-ray flux}. As the binary system rotates, different regions come into view/the line-of-sight. A modulation of the X-ray flux is therefore expected due to changing absorption of the CW emission. In eccentric systems, the variation of the separation $d$ between the stars induce variations in the emitted X-ray flux ($L_X\propto v^{-3.2}d^{-1}$, see Stevens et al. 1992).
\end{itemize}

  \begin{figure}
  \centering
  \caption{$L_X-L_{BOL}$ relation for the NGC6231 cluster. The two colliding wind binaries (circled) clearly lie above the canonical relation. Note that the X-ray emission of \hd326329 is probably contaminated. (From Sana et al.\ 2006a)\label{lxlbol}}
  \end{figure}

For example, in the eccentric binary system \hd93403  (O5.5I+O7V), modulations of the X-ray flux are observed in different energy ranges (see Fig.\ \ref{hd93403} and Rauw et al. 2002). The soft X-ray emission most likely arises in the outer regions of the individual stellar winds and the variability in this energy range is probably associated with opacity effects. In the medium energy band, these effects are much smaller and the observed variation is consistent with a $1/d$ modulation, where $d$ is again the separation between the stars.  \\

  \begin{figure}
  \centering
  \caption{{\it Left:} Schematic view of the \hd93403 binary at the time of the four \xmm\ observations. The top arrows indicate the direction of the observer's line of sight projected on the orbital plane and the dashed and the dotted circles correspond to the surfaces of optical depth unity for the primary wind at 0.5 keV and 1.0 keV respectively. $Right:$ X-ray lightcurve of \hd93403 in the 1.0--2.5 (medium, up), 0.5--1.0 (soft, second from above) and 2.5--10.0 keV (hard, third) energy bands with 1-$\sigma $ error bars. The last panel yields the relative orbital separation between the components of the system while the lower panel provides the position angle of the binary axis ($0^{\circ }$ corresponding to the primary star being ``in front" of the secondary). (From Rauw et al. 2002) \label{hd93403}}
  \end{figure}

  \begin{figure}
  \caption{{\it Left:} Schematic view of \hd152248 at the time of the six \xmm\ pointings. The primary star is in dark grey while the secondary is represented in light grey. Arrows at left-hand indicate the projection, on the orbital plane, of the line-of-sight. $Right:$ Comparison of the observed X-ray luminosities and the results from the hydrodynamical simulations: filled squares represent the dereddened luminosities of the interaction region as predicted by the model; filled circles are the total predicted luminosities (including the expected intrinsic contribution from the two components of \hd152248); open circles show the observed dereddened luminosities. (From Sana et al. 2004) \label{hd152248}}
  \end{figure}

The short-period binary \hd152248 also presents modulations of the X-ray flux that could be reproduced to first order by hydrodynamical modelling (see Fig.\ \ref{hd152248} and Sana et al. 2004).\\

Like these two systems, the binary $\gamma^2$ Vel (WC8+O7.5III) displays phase-locked variations of its X-ray flux. However, these variations are simply due to a changing opacity: when the shock cone around the O-star is in the line-of-sight, the absorption is much smaller than at other phases when the dense wind of the Wolf-Rayet star absorbs most of the X-ray flux (Willis et al. 1995).  \\

Finally, the eclipsing binary CPD-41$^{\circ}$7742 (O9V+B1-1.5V) also presents a modulation of the X-ray flux (see Fig.\ \ref{cpd7742} and Sana et al. 2005). However, this is a peculiar case of CW since the secondary star has a very weak wind. Therefore, it is expected that the wind of the primary directly crashes onto the photosphere of the secondary or that it suffers radiative braking\footnote{In this case, the wind of the primary star is suddenly stopped because of the radiative pressure from its companion. The main consequence of this braking is to push the contact surface away from the secondary star, preventing the wind to actually reach the secondary surface}, leading to a wind-wind interaction very close to the photosphere. Since the interaction region is close to the secondary star, its X-ray emission does not reach us when the secondary star is in front (only the back of the star is then observed) or when the primary occults its companion (Sana et al. 2005).\\

  \begin{figure}
  \caption{{\it Left:} Two schematic views of the geometrical wind-photosphere interaction model in CPD-41$^{\circ}$7742: the (a) panel shows a view from above the orbital plane whereas the (b) panel presents a view along the line of sight just before/after an eclipse. The darker the shading of the secondary surface,  the higher the X-ray luminosity. $Right:$  Predicted unabsorbed flux emitted by the wind-photosphere interaction in CPD-41$^{\circ}$7742 (thick line) compared to the \xmm\ observations (filled triangles are for MOS1, open squares for MOS2). The dashed line gives the intrinsic contribution from the two stellar components. (From Sana et al. 2005) \label{cpd7742}}
  \end{figure}

\subsection{Other Wavelength Ranges}

  \begin{figure}
  \caption{Aperture masking interferometry has enabled to detect a spiral-shape emitting region around the Wolf-Rayet star WR104 (Tuthill et al. 1999). This region corresponds to dust forming in the CW shock and its shape can be explained as a combination of the orbital rotation and the outward motion of the wind (like for a lawn sprinkler). \copyright\ Keck Obs. \label{cwbir}}
  \end{figure}

  \begin{figure}
  \centering
  \caption{VLA maps of the WR147 system (WN8+B0.5V) at 3.6 cm. The asterisks mark the position of the two stars. The northern radio emission is clearly elongated and not centered on the star: it corresponds to the radio emission from the CW region.  (From Contreras \& Rodriguez\ 1999)
\label{cwbradio}}
  \end{figure}

Colliding winds also have an impact on the spectrum at other wavelengths. When binaries composed of certain Wolf-Rayet stars are observed in the infrared (IR), for example, the formation of dust in the CW region can be detected. In eccentric systems, this dust formation is recurrent, since it appears only at specific orbital phases, and it leads to a modulation of the infrared emission (e.g. WR140, Williams et al. 1990). Another example is WR104 (WC9+lateO-earlyB), a CW binary (CWB) surrounded by an IR pinwheel nebula (see Fig.\ \ref{cwbir}) that rotates in harmony with the orbital period.\\

Massive stars generally emit radio waves because of thermal free-free transitions in their winds. Such an emission is of the form $F_{\nu}\propto \nu^{0.6}$. For some binaries, however, an additional, non-thermal emission can be observed. This radio emission is of the synchrotron type and is linked to the motion of relativistic electrons in the stellar magnetic field. The acceleration of electrons to relativistic velocities is probably achieved through the first order Fermi mechanism. Consider a shock in a wind moving with a velocity $V$. A high energy particle crossing the shock hardly notices it. But the downstream gas leaves the shock with a lower velocity ($V/4$) then the velocity at which the upstream gas enters the shock ($V$), and the particle gains on average $0.5 V/c$ by simply crossing the shock downwards. The same particle can then be scattered back upstream by turbulence without any energy loss. A succession of such crossings can thus accelerate the particle to relativistic energies. However, the existence of the pre-existing high energy particles, although necessary, is not yet explained theoretically. On the other hand, it has been proven recently that only shocks between two winds are strong enough to accelerate the particles. This implies that all massive non-thermal emitters are binaries, and this was also proven recently by careful multiwavelength observations (see De Becker, 2005). Note that the radio emission arises in the outer regions of the wind, contrary to what is seen in X-rays, because the inner parts of the winds are opaque to radio waves. The radio emission from CW regions has even been imaged in some cases (see Fig.\ \ref{cwbradio}).\\

Finally, the relativistic electrons can also give rise to non-thermal X- and $\gamma$-rays, in addition to synchrotron radio radiation. In this case, inverse Compton scattering forces the relativistic electrons to give up part of their energy to ambient photons (e.g. the numerous stellar UV photons). These UV photons thus become high-energy radiation. In addition, relativistic protons can produce neutral pions when they interact with the ions in the densest regions of the wind: these pions subsequently decay into $\gamma$-rays. However, this type of emission has not yet been detected  with certainty. \\

\vspace*{1cm}
{\bf \Large Acknowledgements}\\

The author acknowledges the support of the Scientific Cooperation Program 2005-2006 between Italy and the Belgian `Communaut\'e Fran\c caise' (Project 05.02), of the {\sc PRODEX XMM} and Integral projectsand the contracts P5/36 ``P\^ole d'Attraction Interuniversitaire'' (Belgian Federal Science Policy Office). \\

% ===== the bibliography ========

\end{document}